\documentstyle[11pt,moriond_krane,epsfig]{article}

\def\be{\begin{equation}}
\def\ee{\end{equation}}
\def\bea{\begin{eqnarray}}
\def\eea{\end{eqnarray}}

\begin{document}
\vspace*{4cm}
\title{Latest Jet Results From the Tevatron \linebreak
          {\em or QCD: Approaching True Precision} }

\author{ John Krane }

\address{Iowa State University, \linebreak For the D\O\ and CDF Collaborations}

\maketitle\abstracts{
This article summarizes several analyses I discussed 
at Moriond and I attempt to maintain my speaking voice throughout.
I extract several broad themes in the recent development of QCD,
and emphasize how the latest results on jet physics reinforce these
themes: that QCD is in a transitional stage from providing qualitative 
descriptions to providing true quantitative descriptions of the results.  
Certain figures presented in the talk are not
reproduced here in the interest of space, but can be found directly
in the transparencies available online, or in the references to the
refereed journals.
}

\section{Introduction}

I have broken my talk into three parts.  First come the 
experimental details that apply at least generally to all the analyses 
I will discuss.  Second, I have a section describing jet quantities and
event quantities observed at the two experiments.  Finally, I will present
a number of cross section results.

Before I begin, I want to outline what I see as a number of shifting
paradigms in QCD; that is to say, there are shared traditions in our
subfield that must now change.  I believe there are four such trends, 
in both the experimental and theoretical sectors of QCD.

On the theoretical side, a great deal
of recent work has focussed on putting error estimates into the parton
distribution functions (PDFs).  Currently an experimenter performs a 
data-to-theory comparison many times with different PDF sets.  This is certainly
not a valid measure of the uncertainty of the predictions.  Additionally, there has
been some recent progress toward next-to-next-to-leading order calculations
(NNLO) of jet cross sections \cite{glover}.  
On the experimental side, CDF and D\O\ have endeavored to provide
a meaningful estimate of systematic uncertainty, generally in the form of a covariance
matrix.  By quanitfying the correlations of any uncertainty with, for example,
jet energy, it is possible to make $\chi ^{2}$ comparisons between
data and predictions, and actually quantify the level of agreement or disagreement.
Finally, in the future, studies will benefit from much more consistent 
jet definitions.  There is some hope of better underlying event treatment in
data and in Monte Carlo simulations.  Also, there is an ongoing Jet 
Algorithms Workshop \cite{jet_algo} with the two collaborations
trying to eliminate small inconsistencies in their jet algorithms.

All these changes are not iterative so much as 
transformational.  I mention all of this now, not because these 
four themes will play a central role in the remainder of my talk, 
but because they are the background over which I paint these 
experimental results.

\section{Experimental Details}

Most analyses use a simple cone definition of jets, with a radius 
$\sqrt{\left(\Delta \phi\right) + \left(\Delta \eta\right)}=0.7$.  
There is a merging and splitting
decision, where two jets are merged into a single jet if they share
more than a certain fraction of their jet energies.  Theoretical 
predictions require an additional parameter, called $R_{\rm sep}$, to allow
parton-level cones to mimic the merging behavior observed at the 
calorimeter level.

A second algorithm has only recently taken root at Fermilab.  This $k_{T}$
definition of jets \cite{ellis_soper} has the benefit of simultaneous 
validity at the calorimeter level, particle level, and parton level.
The $k_{T}$ is the momentum of one object, projected onto the plane 
that is perpendicular to the other object. 
In essence, the algorithm combines energy clusters into a single entity 
if their relative $k_{T}$ is smaller than some
scaling parameter $D$, where $D=1.0$ or $0.5$ in the present work.  One
may also define subjets by rerunning the algorithm on each jet individually
with a resolution  parameter, $Y_{\rm cut}$, which prevents clusters with 
relatively large separation (but still within $D$) from being merged.

Jet selection criteria at both experiments are largely the same, without 
regard to algorithm.  Without becoming entangled in specifics, suffice 
it to say that it is relatively easy to separate jets from electrons, 
photons, and noise.  The cuts are at least $97\%$ efficient, with a purity
so high that we cannot measure any contamination (estimated at less than
$0.5\%$).

The jet energy correction has three major parts.  First, the offset
correction removes from a jet any noise or energy contributions from the 
underlying event. Second, the calorimeter 
response correction restores the average energy losses due to cracks
in the calorimeters and the losses due to the non-linear energy deposition
of low momentum particles.  Lastly, the showering correction removes
a subtle effect at the cone boundary, where particles that lie inside the 
$R=0.7$ cone prior to hitting the calorimeter, deposit some of 
their energy outside the cone solely because of interactions in the
material of the calorimeter. Collectively, these
three corrections are called the jet energy scale, and further discussion
may be found in D\O\cite{escale} and CDF\cite{escaleCDF} references.

In the case of jet cross sections, or any rapidly-changing distribution, 
fluctuations in jet energy result in a large smearing effect.
The correction is estimated using either the balancing of $p_{T}$ in 
dijet events or through Monte Carlo simulation.

\section{SubJets and Event Quantities}

\begin{figure}[t]
\begin{minipage}{187pt}%\mbox{}
\mbox{\epsfxsize5.4cm\epsffile{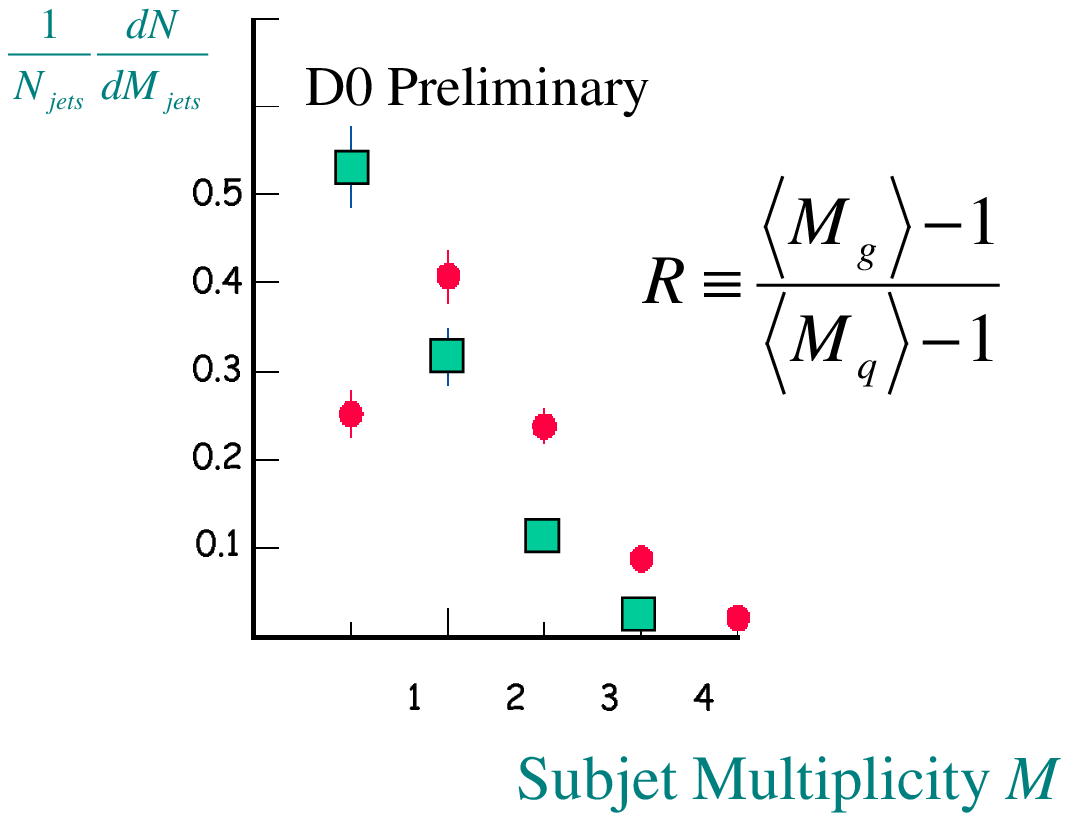}}
\caption{Extracted probabilities of observing $M$ subjets for 
quark-like jets (squares) and gluon-like jets (circles) at 60 GeV.}
\label{subjet}
\end{minipage}
\hspace{14pt}
\begin{minipage}{187pt}%\mbox{}
\mbox{\epsfxsize8.5cm\epsffile{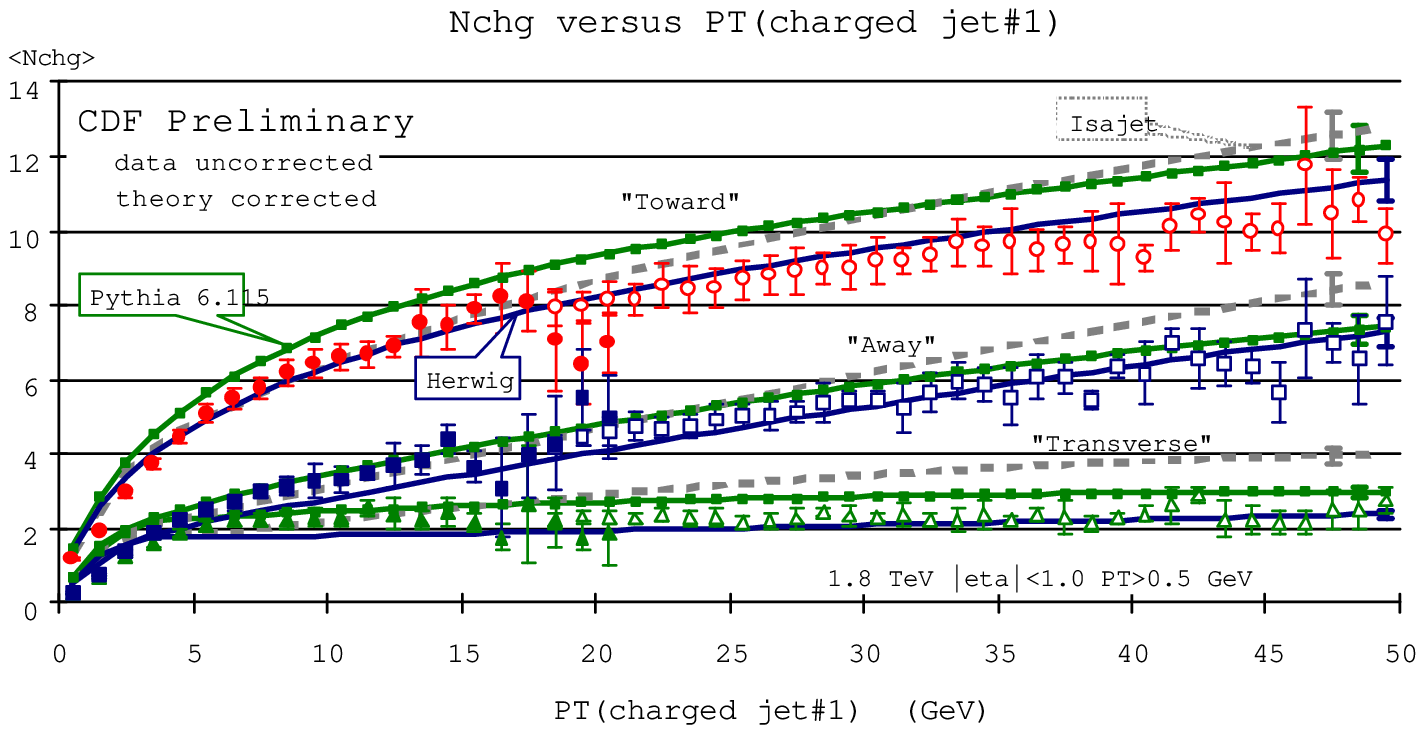}}
\caption{Ability of simulations to model observed charge track distributions.}
\label{pue}
\end{minipage}
\end{figure}

D\O\ uses the $k_{T}$ algorithm to identify
subjets in data collected at its two center-of-mass energies of 630 and 1800 GeV.
Hypothesizing that the multiplicity of subjets for quark parents differs
from that of gluon parents, the expected subjet multiplicity $M$ can be 
expressed as
\[
\left\langle M\right\rangle =f_{g}M_{g}+(1-f_{g})M_{Q}
\]
where $f_{g}$ is the fraction of gluons in the final state provided by PDFs
and Monte Carlo simulation.  Using jets of like energy in both data sets,
and the assumption that $M_{g}$ and $M_{Q}$ depend only on jet energy
(not center-of-mass energy), one extracts the multiplicities characteristic
of the two partons.  Taking the ratio, D\O\ finds
\[
R=\frac{\left\langle M_{g}\right\rangle -1}{\left\langle M_{Q}\right\rangle
-1}=1.91\pm 0.04 \rm{ (stat) }\pm 
\begin{array}{l}
0.23 \\ 
0.19
\end{array}
\rm{ (sys)}
\]
The prediction from Herwig is $R=1.86\pm 0.08 \rm{ (stat)}$.  Figure \ref{subjet} 
provides the spectrum of multiplicities.  Although there has not yet been
any attempt to create a ``gluon likelihood'' function based on these results,
one could imagine significant background reduction in future analyses
if jet parents could be tentatively identified on an individual basis.

At CDF, the central tracker provides a count of charged particles 
as a function of  $p_{T}$ and as a function of position relative to the leading jet.
Within broad sectors, toward, away from, and transverse to the leading jet,
the Monte Carlo generators predict the numbers of particles, with varying
degrees of success.  The simulations are each subdivided into three 
contributions,  which are reweighted to yield the closest 
possible agreement with the data.  Figure \ref{pue} displays the 
data (points) and the reweighted predictions (lines) in the three geometric 
sectors.  The information in this figure is densely packed, exhibiting 
comparisons to all three Monte Carlo
generators simultaneously.   The results 
of this analysis can significantly improve the modeling of the underlying event and 
of low $p_{T}$ physics.

At high transverse energies, NLO QCD is generally in good agreement with collider data.
At low $E_{T}$ however, QCD underpredicts the number of 3-jet and 4-jet
events observed by D\O.  The results, shown in Figure \ref{multijet}, 
debut here at Moriond.  Simple DGLAP treatments suppress events with large
numbers of jets, suggesting a higher order or BFKL-augmented prediction might 
provide better agreement.  Ongoing studies of the angular distributions of these multijet
events should provide an additional handle on the reason for the underprediction
of NLO QCD.

\section{Cross Sections}

At Moriond, I promised and delivered a host of cross section 
results in this section.  During the talk, I showcased several recently published and
recently submitted analyses, but have space to only 
mention them here.  From D\O,
there were three new PRL's:  the so-called R32 analysis, \cite{r32} the ratio of 
dimensionless cross sections, \cite{jet630} and the inclusive jet cross 
section in D\O's full pseudorapidity range. \cite{forward_jet}  Just submitted
to PRD were CDF's central ($0.1< \left| \eta \right| <0.7$) inclusive 
jet cross section \cite{escaleCDF} and D\O's tour-de-force of four jet analyses and
a full description of their covariance matrix techniques. \cite{d0_jet_prd}  
All of these analyses benefit from a rigorous treatment 
of experimental errors that makes possible meaningful $\chi ^{2}$ comparisons.

\begin{figure}[t]
\begin{minipage}{187pt}%\mbox{}
\mbox{\epsfxsize5.8cm\epsffile{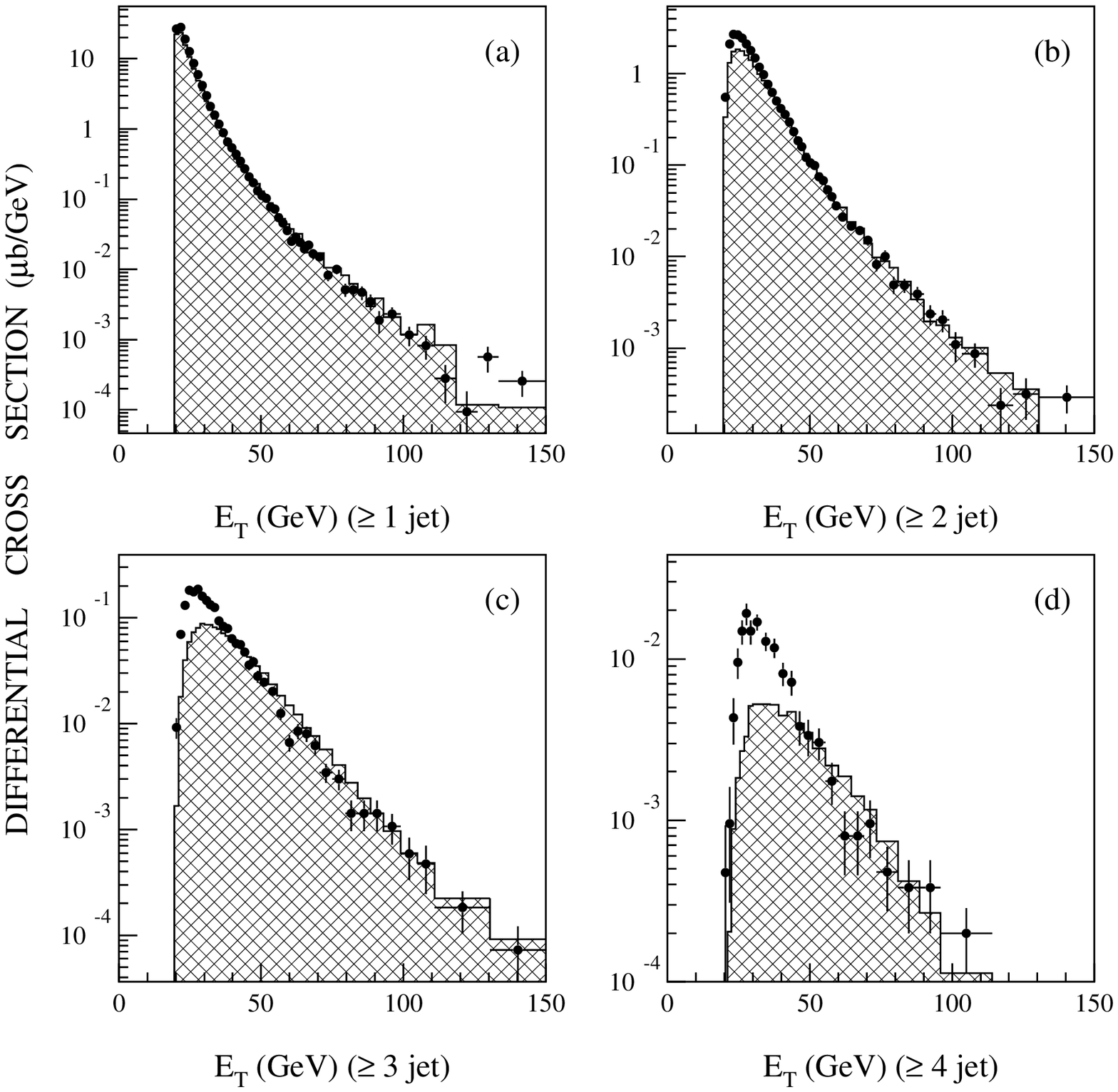}}
\caption{Data (points) and Pythia predictions (hatched) for the $E_{T}$
spectra of njet events.}
\label{multijet}
\end{minipage}
\hspace{14pt}
\begin{minipage}{5.0cm}
      \epsfxsize5.0cm\epsffile{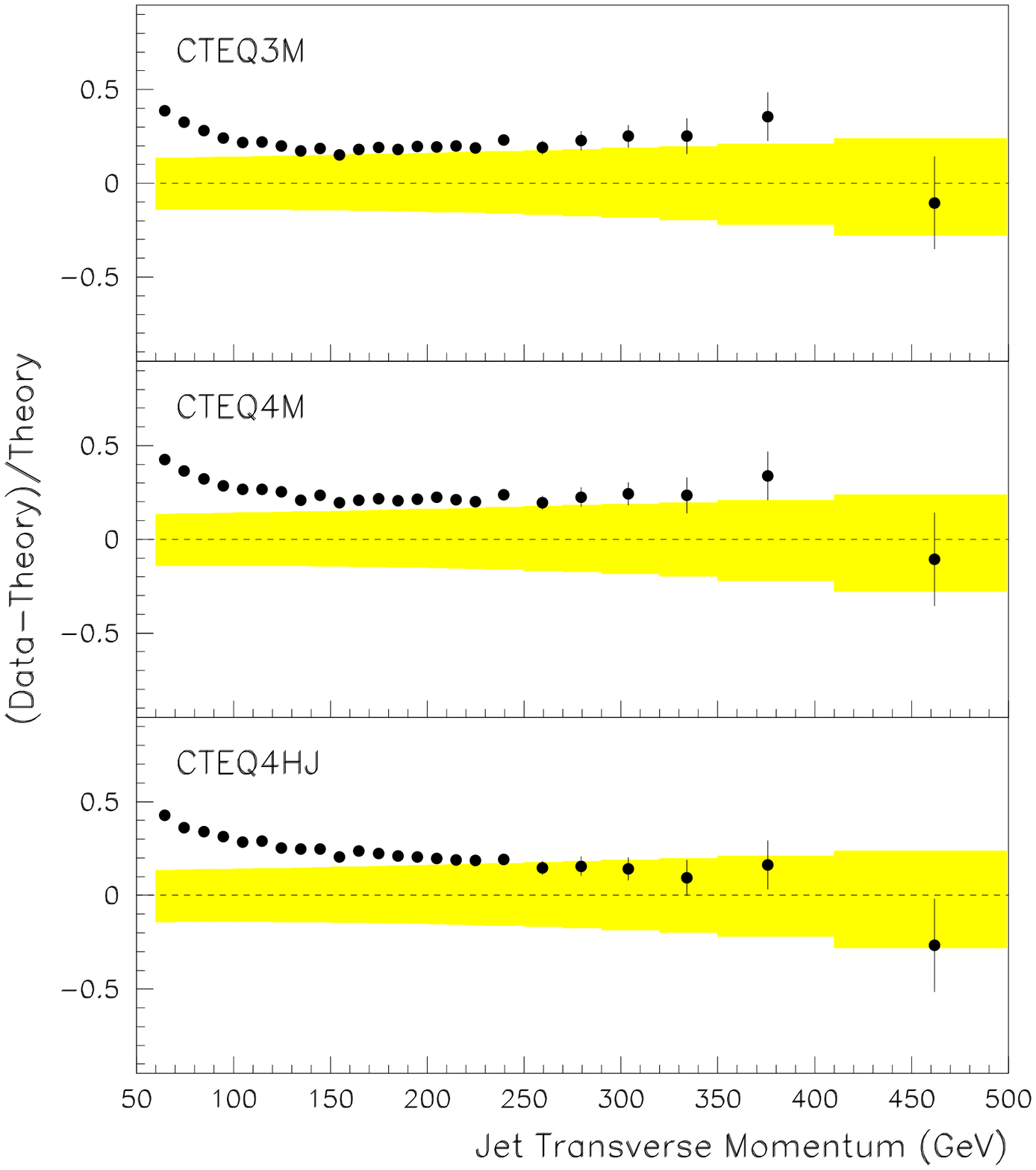}
    \end{minipage}
    \begin{minipage}{3.6cm}
      \caption{Preliminary (Data$-$Theory)/Theory, for $k_{T}$ jet inclusive 
cross section and three NLO QCD predictions.\label{kt}}
    \end{minipage}
\end{figure}

Finally, I present the inclusive $k_{T}$ jet cross section.  This analysis also
makes its debut here at Moriond.  The preliminary results differ from the 
cone-jet analogue by 20\%
or more, with some $p_{T}$ dependence.  D\O\ has an error matrix and
expects to make $\chi ^{2}$ numbers available in the near future.
There are no ``significant'' deviations between data and theory
(Figure \ref{kt}), but 
credibility demands further qualification: the entire distribution
exhibits better than, say, $2\sigma$ agreement, with most deviations 
occuring at low $p_{T}$.  D\O\ is exploring several possible reasons
for the behavior at the low end of the spectrum, which include the 
treatment of underlying event and the effect of final-state hadronization 
on reconstructed energy.

\section {Summary}

The Tevatron stopped running in 1996, but there are still several interesting
analyses in the queue.  I have presented the latest results
on jet substructure, underlying event structure, multijet topology,
and numerous cross section measurements.  I emphasize
the underlying themes of these results, which are
an increased consistency between the jet algorithms of different experiments, 
the extended use of error matrices to quantify results, and continuing work on 
improving the corrections applied to the jet data.  I hope you agree that
these changes, and similar advances in the theoretical sector, are not 
merely incremental improvements;  instead, I believe that
the study of QCD is turning into something quite different than it
was, and within a very short time we will all be talking about 
the latest {\it precision} studies in QCD.


\section*{References}
\begin{thebibliography}{99}
\bibitem{glover} See talk and paper from Nigel Glover, same Moriond session.
\bibitem{jet_algo} The homepage may be found at {\it http://niuhep.physics.niu.edu/~blazey/jet\_alg/jet\_alg.html}.
\bibitem{ellis_soper} The $k_{T}$ algorithm.  Ellis, Soper, Phys. Rev. D48:3160-3166,1993.
\bibitem{escale} The D\O\ energy scale. Nucl.Inst.Meth. A424 (1999) 352-394.
\bibitem{escaleCDF} CDF energy scale and Run 1b jet results, submitted to Phys.Rev.D, hep-ph/0102074.
\bibitem{r32} The ratio, R32.  Phys. Rev. Lett. 86, 1955 (2001).
\bibitem{jet630} Dimensionless cross section ratio.  Phys. Rev. Lett. jet ratio 630/1800.
\bibitem{forward_jet} Rapidity-dependence of the incl. jet cross section.  Phys. Rev. Lett. 86, 1707 (2001).
\bibitem{d0_jet_prd} Multiple D\O\ jet results, accepted by Phys.Rev.D, hep-ex/0012046.

\end{thebibliography}
\end{document}